# Quasiharmonic Lattice Dynamics Calculations of Energy Balances in the Thermal Expansion of ULE® Glass

## James R. Rustad[*]

**Corning Incorporated, Corning NY 14830**
(21 May 2016)

**Abstract-** Quasiharmonic lattice dynamics calculations are carried out on $SiO_2$-$TiO_2$ glass mixtures ($TiO_2 \leq 20$ weight percent) showing that the addition of $TiO_2$ systematically lowers the thermal expansion of high purity fused silica. Adding ten weight percent titania lowers the thermal expansion coefficient by ~0.5 ppm/K in agreement with observations. The driving force for lowering the thermal expansion coefficient in the model system is broken down into mode-by-mode contributions. The driving force is distributed broadly across the frequency spectrum and is not localized into particular frequency ranges that can be associated with the coordination structure of $Ti^{4+}$ in the simulated glass.

## Introduction

Adding a small amount of titanium dioxide (~10 weight percent) to high-purity fused silica (HPFS) lowers the coefficient of thermal expansion (CTE) from approximately 0.5 ppm/K to near zero. This effect (hereafter called the 'ULE effect', with ULE standing for 'ultra low expansion') is the basis for Corning Incorporated's ULE® glass. The underlying reasons behind the ULE effect remain speculative; it has been conjectured that the effect is related to the coordination of $Ti^{4+}$ in the HPFS network, promoting transverse vibrations with negative mode-Grueneisen parameters that counter the usual bond expansion resulting from anharmonicity[1]. A more precise understanding of the mechanism is desired. In this report, lattice dynamics calculations are done on model HPFS and ULE systems to better understand the ULE effect. It is shown that, for sufficiently large systems of ~1500 atoms, lattice dynamics calculations are indeed capable of reproducing something like a ULE effect: although, for the potential model chosen here, the CTE of HPFS is overestimated, adding 10 weight percent $TiO_2$ does lower the CTE by ~0.5 ppm/K, consistent with measurements. The 'mechanism' behind the lowering of the CTE is investigated here by defining, within the quasiharmonic approximation, a driving force for the CTE lowering, and breaking this driving force down into mode-by-mode contributions. If the transverse Ti-O-Si vibrations are the cause of the ULE effect, then contributions at the frequencies of these vibrations should be apparent.

[*] present address University of California, Davis, One Shields Avenue, Davis, CA 95616
(jrrustad@ucdavis.edu)

## Background

In a previous report[2] I calculated the CTE of high-purity fused silica and Ti-bearing fused silica in the quasi-harmonic approximation using the PMMCS empirical potential model[3]. Although the model was able to reproduce some key aspects of the problem, such as the much lower CTE of high purity fused silica relative to $\alpha$-quartz, the addition of $TiO_2$ did not give rise to a definitely identifiable lowering of the CTE for this small system. In fact, for the small systems, in many of the configurations, the addition of ten percent $TiO_2$ raised the CTE. There was an apparent ULE effect predicted by one calculation carried out on a large system (512 $SiO_2$ units), even though the overall CTE for the large system was very close to that calculated for the four small-system configurations. I concluded in Reference 2 that the system size variations were probably within calculation uncertainties having to do with the inherent conformational variability in "virtual system preparation" through the simulated annealing cycles. Configurations produced by the simulated annealing procedure differ from one another, resulting in variations in the CTE. I had thought the single calculation showing a ULE effect for the large system was just noise. This report is a follow-up to the previous study, with a much more systematic examination of the CTE-lowering effect on large systems. As the present study will show, there really does appear to be a size-dependence, not in the overall predicted CTE, but in the predicted CTE lowering by addition of $TiO_2$. This size effect arises largely from resolution– the small systems exhibit more configuration-to-configuration scatter and can't resolve a consistent CTE-lowering effect associated with $TiO_2$ addition.

## Methods

In the harmonic approximation, the free energy $F$ can be written as[4]:

$$F(V,T) = U(V) + F_{vib}(V,T) \qquad (1)$$

where $F_{vib}(V,T)$ is the vibrational contribution to the free energy:

$$F_{vib}(V,T) = \sum_i \tfrac{1}{2}hc\omega_i(V) + k_b T \ln\left(1 - e^{-hc\omega_i(V)/k_b T}\right) \qquad (2)$$

where $i$ runs over the 3N frequencies, $V$ is volume, $T$ is temperature, $\omega$ is frequency, and the "$\omega_i(V)$" (i.e. "frequency $i$ evaluated at volume $V$") is a reminder that the "quasiharmonic" vibrational frequencies are, due to anharmonicity, implicit functions of the volume. Larger volumes generally imply lower harmonic vibrational frequencies and, therefore, lower $F_{vib}(V,T)$. The term "quasiharmonic" means that the frequencies are computed in the harmonic approximation at a given fixed volume; but that the calculated harmonic frequencies are functions of the imposed system volume. $U(V)$ is given, in the pair potential model used here, by the following equation:

$$U(V) = \sum_i \sum_{j'} \frac{z_i z_j}{r_{ij}} + D_{ij}\left[\left\{1 - e^{-a_{ij}(r_{ij}-r_{0ij})}\right\}^2 - 1\right] + \frac{C_{ij}}{r_{ij}^{12}} \quad (3)$$

where the first sum is over all atoms $i$, the second is the sum over atoms $j$ with the prime on $j$ indicating that $j \neq i$. The parameters of this equation (the effective charges $z_i$, as well as $D_{ij}$, $a_{ij}$, $r_{0ij}$, and $C_{ij}$) are given for Si-O, Ti-O, and O-O interactions in Reference 3, where they have been fitted to the structure and elastic constants of α-quartz (Si-O, O-O) and rutile (Ti-O). Si-Si, Si-Ti, and Ti-Ti interactions have $D_{ij}$, $a_{ij}$, $r0_{ij}$, and $C_{ij}$ equal to zero (cation-cation repulsion is expressed only through the effective charge).

Four systems of 512 $SiO_2$ formula units, and four systems with 472 $SiO_2$ formula units and 40 $TiO_2$ formula units (10.1 wt % $TiO_2$), were "annealed" from 6000 K to 300 K over $2 \times 10^8$ time steps (time step = 0.27 fs). Three of the four systems were annealed at a constant volume corresponding to a density of 2.2 g/cm$^3$ for pure fused silica. The cell size in these calculations was 2.85 nm. One system was annealed at zero pressure (giving a much larger density of 2.6 g/cm$^3$). In addition, one of the low-density systems was prepared with 432 $SiO_2$ and 80 $TiO_2$ formula units (20 wt. % $TiO_2$). In this system, the $TiO_2$ concentration far exceeds the solubility in real fused silica. This is possible in molecular dynamics calculations because the very fast quenching rates and small system sizes inhibit nucleation of solid phases. The high-concentration runs are used to ensure that the magnitude of the ULE effect is proportional to the $TiO_2$ content and to help establish that the calculated CTE-lowering is really due to the added $TiO_2$.

After annealing, the lattice parameters (a,b,c,α,β,γ) and fractional coordinates of each system were optimized to minimize the lattice energy U at zero temperature. Each of these zero-temperature systems was then subjected to free energy minimization in the quasiharmonic approximation according to Equation 1 over a range of temperatures from 0 to 450K in increments of 50K using the GULP code[5]. The zero static internal stress approximation (ZSISA)[6] was employed in the free energy minimization as has been recommended for silica polymorphs[7].

## Results

The calculated CTE values for all four 2.85 nm systems having 10 weight percent $TiO_2$, as well as the 20 weight percent $TiO_2$ system, are shown in Figure 1. Overall, the CTE is overestimated with the PMMCS model. To put the estimate in context, Figure 2 shows the CTE computed for a 192-atom system of fused silica (configuration taken from Reference 2) computed using the SLC[8] shell-model, as well as measured values from push-rod dilatometry[9]. Figure 2 shows that the absolute CTE predicted for fused silica is highly model-dependent. Both potential functions predict a strong CTE lowering in going from α-quartz to fused silica, but precisely how close the CTE comes to zero depends on the accidents of model parameterization. This probably indicates that parameterized models (at least models that were fitted to something other than the CTE or phonon spectra) are not up to the task of predicting CTE accurately, and that first-principles calculations should be used. Because this study and Reference 2 indicate that the lowering effect is only seen consistently for the 2.85 nm system, this would mean doing first-principles free-energy minimization calculations on ~1500 atom systems, which is currently on, if not over, the edge of feasibility even with substantial computational resources.

However, despite the overall overestimation of the CTE, the lowering of the CTE by addition of $TiO_2$ (i) is consistent over all configurations tested, (ii) systematically increases with increasing $TiO_2$ content, and (iii) is close to that observed experimentally. It is therefore worthwhile to look in more detail at the underlying causes of the reduction in CTE in the model system.

Consider some elementary ideas behind the driving force for CTE in the context of the quasiharmonic approximation. It will be convenient to denote the lattice energy at zero temperature, without the vibrational contribution, as $U_0$, and the volume as $V_0$. First, whether or not the material exhibits positive or negative CTE, $\Delta U = U(V,T)-U_0$ is certainly positive as, by definition, $U_0$ corresponds to the atomic coordinates that minimize $U$. If the system changes its volume at temperature T, the vibrational contribution to the free energy $F_{vib}(V_T,T)-F_{vib}(V_0,T)$ must be negative and must more than offset the energy $\Delta U$ required to expand the lattice:

$$[U(V,T) - U_0] + [F_{vib}(V_T,T) - F_{vib}(V_0,T)] < 0 \tag{4}$$

Note carefully that the second term is a difference in $F_{vib}$ between two states at the same temperature $T$, but one having the expanded volume $V_T$ and one having the unexpanded volume $V_0$. This energy difference could be thought of as a sort of potential energy or "driving force" for thermal expansion. The second term in Equation 4 can be written as a sum over contributions from individual vibrational modes:

$$\Delta F_{vib}(T) = F_{vib}(V,T) - F_{vib}(V_0,T)$$
$$= \sum_i \tfrac{1}{2} hc[\omega_i(V) - \omega_i(V_0)] + k_b T \left[\ln\left(1 - e^{-\frac{hc\omega_i(V)}{k_b T}}\right) - \ln\left(1 - e^{-\frac{hc\omega_i(V_0)}{k_b T}}\right)\right] \tag{5}$$

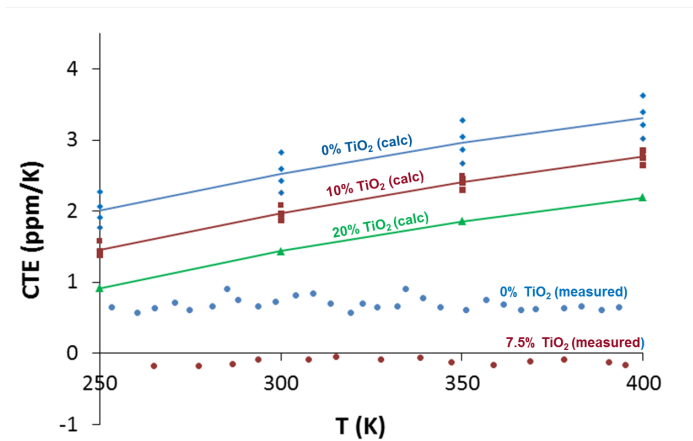

**Figure 1.** CTE versus T for fused silica with varying amounts of added $TiO_2$.

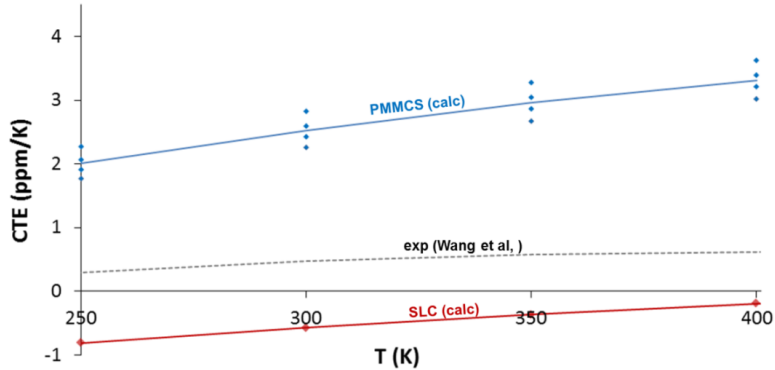

**Figure 2.** CTE vs. T for high-purity fused silica, calculated with two different potential models. Blue line is the average for the CTE calculated from the Pedone et al (2006) (PMMCS) potential, red line is the result of a calculation using the shell model potential of Sanders et al (1984) (SLC). Experimental measurement is taken from Reference 9.

It is then interesting to see which frequencies make the greatest contribution to the driving force for thermal expansion, or conversely, in the ULE, to see how Ti substitution reduces this driving force, and whether this reduction is concentrated at particular frequencies, such as, for example Ti-O-Si transverse vibrational modes, as has been speculated in Reference 1.

**Table 1.** Volume and energetic quantities for expansion from 300 K to 400 K.

|  | $\Delta V$ Å$^3$ | $\Delta U$ (eV) | $\Delta zpe$ (eV) | $\Delta \ln$ (eV) | $\Delta F_{vib}$ (eV) | $\Delta U + \Delta F$ (eV) |
|---|---|---|---|---|---|---|
| HPFS |  |  |  |  |  |  |
| 1 | 15.333 | 0.058480 | -0.050861 | -0.010441 | -0.061302 | -0.002822 |
| 2 | 17.915 | 0.046875 | -0.040081 | -0.009298 | -0.049379 | -0.002504 |
| 3 | 19.106 | 0.047994 | -0.041159 | -0.009363 | -0.050522 | -0.002528 |
| 4 | 21.481 | 0.048857 | -0.040915 | -0.010763 | -0.051678 | -0.002821 |
| *avg.* | *19.501* | *0.047909* | *-0.040718* | *-0.009808* | *-0.050526* |  |
| ULE |  |  |  |  |  |  |
| 1 | 13.440 | 0.051587 | -0.046395 | -0.007483 | -0.053878 | -0.002290 |
| 2 | 16.052 | 0.039444 | -0.035185 | -0.006137 | -0.041322 | -0.001878 |
| 3 | 15.499 | 0.037574 | -0.033904 | -0.005408 | -0.039312 | -0.001738 |
| 4 | 16.104 | 0.036780 | -0.032509 | -0.005903 | -0.038412 | -0.001632 |
| *avg.* | *15.885* | *0.037933* | *-0.033866* | *-0.005816* | *-0.039682* |  |

**Table 2.** Bulk and shear modulus of glass at a volume corresponding to 400 K.

|  | $\rho$ (g/cm$^3$, 400 K) | K (GPa) | $\mu$ (GPa) |
|---|---|---|---|
| HPFS |  |  |  |
| 1 | 2.677 | 72.4 | 47.1 |
| 2 | 2.450 | 45.9 | 35.0 |
| 3 | 2.439 | 43.7 | 34.7 |
| 4 | 2.336 | 37.9 | 31.5 |
| ULE |  |  |  |
| 1 | 2.689 | 77.1 | 44.5 |
| 2 | 2.437 | 45.2 | 34.1 |
| 3 | 2.435 | 45.2 | 34.0 |
| 4 | 2.394 | 41.1 | 30.2 |

Table 1 shows the data for the four trials of HPFS and ULE, giving the volume expansion from T=300 to T=400, the (unfavorable) $\Delta U$ in lattice energy required to achieve the observed expansion, the (favorable) $\Delta F_{vib}$ offsetting the $\Delta U$, as well as the total change in free energy in the last column. The $\Delta F_{vib}$ is broken down into contributions from the zero point energy and logarithmic term in Equation 5. The averages of each entry of trials 2-4 (each of these annealed at the experimental glass density) are given in bold italics. It is important to remember that the quantities reported in Table 1 are not normalized in any way, and represent "raw" quantities for the 512 $SiO_2$ and 472 $SiO_2$/40 $TiO_2$ systems between 300K and 400K. Table 1 clearly shows the reduced thermal expansion in the case of the ULE glass. It is also apparent that, for HPFS, ~80 percent of the driving force for the thermal expansion comes from the zero-point energy reduction upon volume expansion, with ~20 percent coming from the logarithmic term. In the case of ULE, about 85 percent of the driving force for expansion comes from the zero-point energy contributions. The annealing density has a noticeable effect, with a larger driving force for expansion at high density, but somewhat less total expansion. The reason for the lower expansion must certainly be the higher modulus of the denser glass (see Table 2). Note that the densities in Table 2 are not equal to the annealing densities, as the glass compacts somewhat on the quenching to zero temperature before the quasiharmonic calculations are started.

Figure 3 shows the breakdown of the individual contributions to $\Delta F_{vib}$ as a function of frequency (i.e. $\Delta F_{vib}(\omega_i)$ vs. $\omega_i$). To clarify, the sum of all the contributions shown in each panel of the figure would be equal to the $\Delta F_{vib}$ entry in Table 2. The contributions are also broken down into separate contributions from the zero-point energy and the logarithmic term in Equation 5. To make it easier to see the total contribution, the cumulative sum of the contribution to $\Delta F_{vib}(\omega_i)$, for the thee systems annealed at low density, is given in Figure 4.

The striking aspect of Figures 3 and 4 is the uniformity of the spectral contributions to the CTE-lowering effect. There is no single region of the frequency spectrum where such contributions are made. For the logarithmic term (accounting for 15-20 percent of the driving force for expansion) most of the contributions (positive or negative) are made in the low-frequency region (less than 400 $cm^{-1}$), and there is a clear but broad tendency for the ULE to lie above the HPFS. For the zero-point-energy term (accounting for 80-85 percent of the driving force for expansion) the largest contributions are made above 600 $cm^{-1}$, and, everywhere, the curve representing the ULE contributions lies above the HPFS. The noisy region near 900 $cm^{-1}$ corresponds to a gap in the vibrational density of states. Because of the simple form of the zero-point energy contribution, it is easy to see that the addition of $TiO_2$ to the network decreases the frequency shift on expansion (the mode-Grueneisen parameter) of nearly every normal mode in the system, and, as shown in Figure 1, this uniform frequency shift is sufficient to lower the thermal expansion to the extent that is observed in nature.

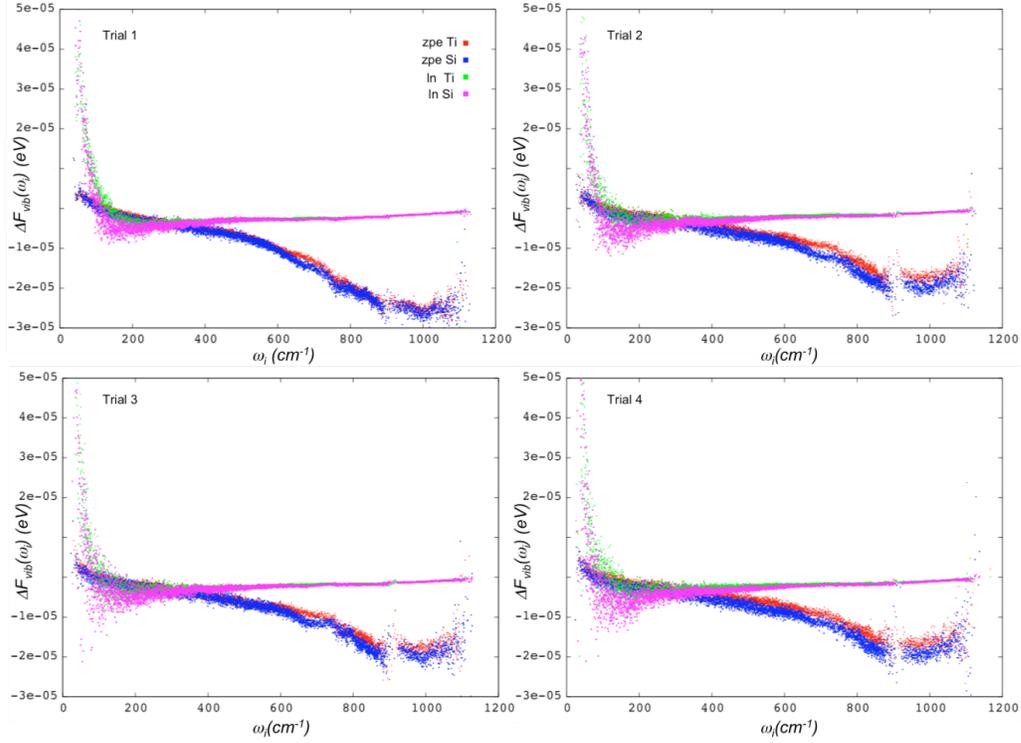

**Figure 3.** Breakdown of mode-by-mode contributions to $\Delta F_{vib}$ (between 300K and 400K). Magenta (HPFS) and green (ULE) dots represent contributions from the logarithmic term, while blue (HPFS) and red (ULE) dots represent contributions from the zero-point energy term.

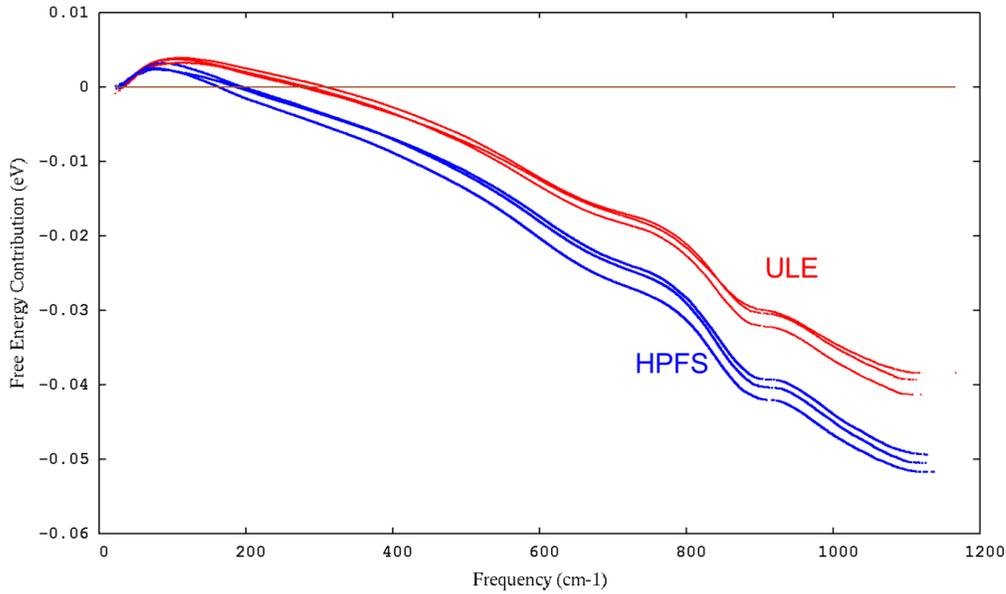

**Figure 4.** Cumulative contributions to $\Delta F_{vib}$ (from 300K to 400 K) for the three systems annealed at low density for HPFS (blue) and ULE (red) compositions.

**Discussion**

These results raise several questions about mechanisms. In this context, it is important to point out that $TiO_2$ is unique in the magnitude of the ULE effect. None of the other tetravalent oxide additives (e.g. $GeO_2$, $ZrO_2$, $SnO_2$) give the desired lowering of CTE. This would seem to indicate that there is something special about the structure of $Ti^{4+}$ in HPFS that 'causes' the ULE effect, and this fact no doubt led to much of the speculation surrounding the mechanisms of CTE-lowering and how it might relate to the specific structural characteristics of $Ti^{4+}$ in HPFS[1]. The results presented in this report are, in this respect, somewhat unsatisfying. The behavior of the model system, which gives contributions more or less continuously across all frequencies, seems to be hard to relate to any unique structural aspects of ULE glass. If the frequency contributions to the ULE effect are spread so blandly and uniformly across the spectrum, why is it that only $Ti^{4+}$ would act as a CTE suppressant?

It is worth pointing out that model ULE structures generated with the PMMCS potential do, in fact, show a range of $Ti^{4+}$ coordination numbers. A more complete analysis is in progress and will be given in a subsequent report, but there are significant numbers of five-fold coordinated $Ti^{4+}$ in the generated structures, that is, the model ULE structures generated by the PMMCS force field do have some of the structural features that have been speculated to give rise to the lowering of the CTE in ULE glass, even though the contributions to the CTE lowering cannot be easily assigned to these structural features (and even though these high-coordinated $Ti^{4+}$ may not actually be present at ULE Ti concentrations for glasses processed at high temperature; see Reference 10).

Vibrational modes with negative mode-Grueneisen parameters (i.e. the frequency increases with increasing volume), of course, suppress thermal expansion. In earlier work, it had been speculated that there would be a band of negative mode-Grueneisen parameters associated with transverse Ti-O-Si vibrations that gave rise to the ULE effect. Figure 3 shows that the negative mode-Grueneisen parameters are certainly present but make contributions only at low frequencies (less than 150 cm$^{-1}$). They are present both in the HPFS and ULE glasses (it is these modes that give HPFS a much lower CTE than $\alpha$-quartz), and are not unique to ULE in any way. Overall, they do make slightly less positive contributions for the HPFS than the ULE, but are by no means the 'cause' of the ULE effect in the model system. As easily seen in Figure 4, even more contribution is made by the difference in the zero-point energy terms across the higher frequency range.

Figure 3, in essence, obviates the necessity of having to imagine that there must be localized phenomena such as double-well potentials associated with rigid-unit motions, or with anomalous coordination of Ti (e.g. as O---$TiO_5$) or peculiar transverse Ti-O-Si bonds that lie behind the role of $TiO_2$ in ULE glass, such as discussed in Reference 1

and elsewhere. While these are all interesting ideas, and might certainly contribute to thermal expansion in real ULE glass, they are not *necessary* to understand, in a general sense, the CTE-lowering effect of $TiO_2$, as we have here a model system which clearly and consistently shows the CTE-lowering that doesn't result from any of these effects (it is worth emphasizing that there are no double-well phenomena taken into account in the quasiharmonic calculations carried out here; each of the configurations of both HPFS and ULE have been optimized to a single local minimum). The predicted ULE effect, in the context of this model, is a structurally delocalized phenomenon. The puzzle is then to reconcile the uniqueness of $TiO_2$ in the context of such a collective mechanism. But the key result of these calculations, that there at least *can be* a collective response to $TiO_2$ addition that pervasively lowers the modal contributions to CTE suppression across the frequency spectrum, is unexpected and useful for thinking further about possible mechanisms.

---

[1] Schultz PC and Smyth HT (1970) Ultra-low expansion glasses and their structure in the $SiO_2$-$TiO_2$ system. Amorphous Materials, eds. Douglas R.W. and Ellis B., Wiley Interscience New York 453.

[2] Rustad JR (2014) Lattice dynamics calculation of the thermal expansion of pure and Ti-bearing fused silica in the quasi-harmonic approximation arXiv:1404.0264 [cond-mat.mtrl-sci].

[3] Pedone A, Malavasi G, Cristina MC, Cormack AN, Segre U (2006) A new self-consistent empirical interatomic potential model for oxides, silicates, and silica-based glasses. J. Phys. Chem. B, **110**, 11780-11795.

[4] Baroni S, Giannozzi P, Isaev E Thermal properties of materials from ab initio quasi-harmonic phonons. http://arxiv.org/abs/1112.4977.

[5] Gale JD (1997) GULP - a computer program for the symmetry adapted simulation of solids, *JCS Faraday Trans.,* **93**, 629.

[6] Allan NL, Barron TKH, Bruno JAO (1996) The zero static internal stress approximation in lattice dynamics, and the calculation of isotope effects on molar volumes. J. Chem. Phys., 105:8300-8303.

[7] Gale JD (1998) Analytical free energy minimization of silica polymorphs. J. Phys. Chem. B, 102, 5423-5431.

[8] Sanders MJ, Leslie M, Catlow CRA (1984) Interatomic potentials for $SiO_2$. *J. Chem Soc. Chem. Commun* 1984,1271-1273.

[9] Wang H, Yamada N, Okaji M, Proceedings of the Thirteenth International Thermal Expansion Symposium, Technomic Publishing Company, Lancaster, PA, 2000.

[10] Pickup DM, Sowry FE, Newport RJ, Gunawidjaja PN, Drake KO, Smith ME (2004) The Structure of $TiO_2$−$SiO_2$ Sol−Gel Glasses from Neutron Diffraction with Isotopic Substitution of Titanium and $^{17}O$ and $^{49}Ti$ Solid-State NMR with Isotopic Enrichment. *J. Phys. Chem. B*, 108, 10872-10880.